\documentclass[12pt]{article}

\usepackage{amsmath, amssymb, amsthm}
\usepackage{bm, bbm}

\usepackage{graphicx}
\usepackage{float}
\usepackage{multirow}
\usepackage{subcaption}
\usepackage{rotating}
\usepackage{array}

\usepackage{setspace}
\usepackage{fullpage}
\usepackage{parskip}

\usepackage{algpseudocode}

\usepackage{comment}
\usepackage{url}
\usepackage[normalem]{ulem}

\usepackage{natbib}

\usepackage[dvipsnames]{xcolor} % for named colors like SkyBlue, BrightRed
\usepackage{hyperref}
\hypersetup{
  pdfborder = {0 0 0.5 [3 3]},
  colorlinks = true,
  linkcolor  = BrightRed,
  citecolor  = SkyBlue
}

\newcolumntype{L}[1]{>{\raggedright\let\newline\\\arraybackslash\hspace{0pt}}m{#1}}
\newcolumntype{C}[1]{>{\centering\let\newline\\\arraybackslash\hspace{0pt}}m{#1}}
\newcolumntype{R}[1]{>{\raggedleft\let\newline\\\arraybackslash\hspace{0pt}}m{#1}}

\newcommand{\by}{\bm{y}}
\newcommand{\bx}{\bm{x}}
\newcommand{\bbeta}{\boldsymbol{\beta}}

\definecolor{SkyBlue}{RGB}{14, 118, 188}
\definecolor{BrightRed}{RGB}{223, 82, 78}
\definecolor{Green638}{RGB}{165,255,118} 
\definecolor{FoamGreen}{RGB}{25, 200, 150}

\title{Bayesian Multinomial Logistic Regression for \\Numerous Categories}

\author{\textbf{Jared D. Fisher}\thanks{Contact us at fisher@stat.byu.edu and/or kylemcevoy@ucla.edu. We thank Li (Kelly) Kang, David W. Puelz, and Carlos M. Carvalho for their early work on this project. We also thank Candace Berrett for her helpful comments. This work was partially supported by the U.S. National Science Foundation DMS RTG Grant \#1745640.} \\ 
{\normalsize Brigham Young University} \and \textbf{Kyle R. McEvoy}\\ 
{\normalsize University of California, Los Angeles}
}

\date{\today}

\begin{document}
 
\maketitle
 
\begin{abstract}
 Bayesian multinomial logistic regression provides a principled, interpretable approach to multiclass classification, but posterior sampling becomes increasingly expensive as the model dimension grows. Prior work has studied scalability in the number of subjects and covariates; in contrast, this paper focuses on how computation changes as the number of outcome categories increases. 
 To improve scalability in settings with numerous categories, we adapt a gamma-augmentation strategy to decouple category-specific coefficient updates, so that each category’s coefficients can be updated conditional on a single auxiliary variable per subject, rather than on the full set of other categories’ coefficients. Because the resulting coefficient conditionals are non-conjugate, we couple this augmentation with either adaptive Metropolis-Hastings or elliptical slice sampling. 
 Through simulation and a real-data example, we compare effective sample size and effective sampling rate across several standard competitors. We find that the best-performing sampler depends on the dimension and imbalance regime, and that the proposed augmentation provides substantial speedups in scenarios with numerous categories.   
\end{abstract}


\section{Introduction}
\label{sec:intro}

Standard logistic regression is one of the most popular approaches to probabilistic binary classification: the problem of assigning a probability of being in one of two categories to an observation. Compared to popular black-box machine learning approaches used for classification, logistic regression has the added bonus of interpretability: we can clearly state, or even plot, the model's functional relationship between any input variable and the probability of interest. This is possible as the core of the model is additive, namely a linear model of the log odds.

When considering more than two categories, the natural extension is 
multinomial logistic regression (MLR).
Bayesian approaches to coefficient estimation in multinomial logistic regression naturally yield uncertainty quantification but may require more computation time than other common methods, as the posterior distribution of the coefficients is not in a convenient analytical form. Thus, in order to estimate these coefficients' posterior distributions, it is often necessary to pursue Markov chain Monte Carlo (MCMC) strategies that can be computationally intensive. 

While Gibbs methods have historically been preferred to Metropolis-Hastings samplers in most MCMC applications, standard MLR models do not allow sampling from closed-form conditional distributions.  Thus, handling computational challenges in posterior sampling for MLR is often approached through data augmentation techniques, which, by adding auxiliary variables to the model, allow for samples to be drawn directly from posterior conditional distributions with Gibbs sampling instead of using Metropolis-Hastings or other techniques. This can actually improve the ability of MCMC to efficiently estimate posterior distributions, even though the parameter space has increased.   \cite{holmes2006} present a MCMC sampling strategy that conditions on two sets of carefully-chosen auxiliary variables and the data such that the coefficients' prior and posterior both follow multivariate normal distributions. \cite{FSF2010} and \cite{FSF2012} find that their alternative setup using a single set of auxiliary variables with a difference random utility model yields better computation time and effective sample size than \cite{holmes2006}. \cite{Windle2013} introduce the P\'olya-Gamma data augmentation strategy, and further find improved performance over \cite{FSF2010}. P\'olya-Gamma augmentation has since become the go-to approach for Bayesian logistic regression, with over 1500 citations on Google Scholar as of January 2026, both for logistic regression and for embedding logistic regression within larger Bayesian models. 

Some recent work has looked at the samplers' performance on categorical response variables that are imbalanced, e.g. when there are many cases of $y_i = 0$ but few cases of $y_i = 1$. \cite{johndrow2019mcmc} find that when observations are imbalanced across categories, adaptive Metropolis-Hastings samplers outperform P\'olya-Gamma data augmentation in terms of sampling efficiency. \cite{zens2023ultimate} then present the Ultimate P\'olya-Gamma (UPG) sampler, which uses two levels of data augmentation to the traditional P\'olya-Gamma sampler's one. 
They show that UPG achieves similar sampling efficiency to the adaptive Metropolis method of \cite{johndrow2019mcmc}, though each method has different strengths. UPG sampling performs better when categories are relatively balanced and/or the sample size is small, while adaptive Metropolis-Hastings sampling performs better when class imbalance is strong.
 
In this paper, we ask a different question: how do methods perform when there are many categories of interest? While all of the aforementioned papers present sampling schemes for MLR, the focus is typically on logistic regression for binary classification and extensions to multinomial classification applications are often presented with only three categories. The two exceptions that we have found came in technical supplements. First, the supplement of \cite{Windle2013} looks at their method's classification accuracy on a dataset with six categories. Second, \cite{johndrow2019mcmc} compares the sampling efficiency of adaptive metropolis for logistic versus probit regression as the sample size increases in simulated categorical data with severe imbalance across four classes. Hence, the computational efficiency of multinomial logistic regression methods is under-studied when outcomes have more than a small handful of categories. 
 
We present two contributions. First, we propose an alternative data augmentation technique that is a specialization of the gamma-augmentation strategy in \cite{Murray} to the standard linear-predictor MLR setting. The novelty of this augmentation for MLR 
is that, given a single auxiliary variable for each subject, the posterior sampling of one category's regression coefficients does not depend on the coefficient parameters from other categories. This decoupling of categories yields simpler sampling of the regression coefficients using standard sampling techniques. 
In this way, the method can handle larger numbers of categories with less reduction in computational efficiency compared to other methods by employing advantages of both traditional samplers and data augmentation techniques.  Our approach works for most standard prior distributions on the regression coefficients, and though the posterior distributions on the covariates are not standard distributions allowing full Gibbs updates, we find that both Metropolis-Hastings and elliptical slice samplers yield high effective sampling rates. 

Second, we examine the computational efficiency of sampling algorithms under various simulation settings. As previous work in this area typically only looks at categorical outcomes with three categories, our analysis presents situations where categorical outcomes have up to 100 categories.  While the aforementioned methods are indeed good approaches when there are a small number of categories, our simulation studies show that in many scenarios when categorical response variables with large number of categories, our data-augmentation methods yield improved sampling efficiency of the MLR model coefficients. 

The remainder of this article proceeds as follows. Section \ref{sec:meth} presents our proposed sampling method. Section \ref{sec:sim} discusses our simulation studies. 
Section \ref{sec:data} shows how these MNL methods work on a real dataset. 
Section \ref{sec:discussion} concludes.

\section{Methodology}
\label{sec:meth}
For  $i \in \{1, \ldots, N\}$, let $\by_i$ follow a multinomial distribution,
\begin{align}
    \by_i \rvert \cdot \sim \mathrm{Multinomial}(n_i, \pi_{i1}, \ldots, \pi_{iC}),  
\end{align}
consisting of $n_i$ independent categorical observations from subject $i$, where each observation comes from category $j$ with probability $\pi_{ij}$, $j \in \{1, \ldots, C\}$.
Logistic regression is used to model each of the categorical probabilities. That is, let $\bbeta_j \in \mathbb{R}^P$ be a vector of coefficients for category $j$, such that the probability that an observation from $\by_i$ falls into one each of the $C$ different categories is modeled with a softmax function the linear predictors $\bx_i^T\bbeta_k$, $k=1,...,C$: 
\begin{align}
    \pi_{ij} = \pi_j(\bx_i) = \frac{\exp{ \bx_i^T \bbeta_j }}{\sum_{k = 1}^C \exp{\bx_i^T \bbeta_k}   }
\end{align}
where each vector $\bx_i \in \mathbb{R}^P$ contains the explanatory variables associated with subject $i$.

In this paper, we assume 
that the joint prior on the coefficient vectors $\bbeta_j$ factors into a product of the marginal priors for each category. This should include most commonly-assumed priors, and independence across categories is sufficient for such factorization of priors.  This assumption yields the following posterior on the vector of coefficients $\bbeta_j$:
\begin{equation}
    p(\bbeta_j|\cdot) 
    \propto \left[\prod_{i=1}^N   \frac{ \exp \left\{  y_{ij} \bx_i^T \bbeta_j \right\}   }{\left(\sum_{k = 1}^C \exp \{\bx_i^T \bbeta_k\} \right)^{n_i }   }\right]  p(\bbeta_j). 
\end{equation} 
Note that the denominator $\sum_{k = 1}^C \exp \{\bx_i^T \bbeta_k\}$ implies that the posterior of $\bbeta_j$ is conditional upon the coefficients $\bbeta_k$ from all other categories $k \ne j$. Thus the calculation of the sum in the denominator above grows in complexity as $C$ grows.  We also assume that the final category is a baseline, i.e. $\bbeta_C = 0$, so that we can identify the coefficients $\bbeta_j$, $j=1,...,C-1$.

\subsection{Data Augmentation}

Standard multinomial logistic regression, such as used in the papers discussed in Section \ref{sec:intro}, is formulated such that information in covariates $\bx_i$ enter into the model as parametric linear combinations. 
However, these could be any category-specific function of the data, $f_j(\bx_i)$, such that
\begin{align}\label{pi_f}
    \pi_{ij} = \pi_j(\bx_i) = \frac{f_j( \bx_i)}{\sum_{k = 1}^C f_k(\bx_i)}.
\end{align} This is the case in \cite{Murray}, who fits these functions with Bayesian additive regression trees \citep{BART}. 
However, the denominator above presents a challenge with MCMC: the conditional posterior of $f_j$ depends on $f_k$ for all $k\ne j$.
To solve this problem, \cite{Murray} augments the data with the auxiliary variables $\phi_i$, where
\begin{align}
    (\phi_i|\cdot) \sim \text{Gamma}\left(n_i, \sum_{j=1}^C f_j(x_i)\right),
\end{align}
akin to the ideas in \cite{Nieto2004} and \cite{Walker2011}. 
This yields the joint probability model 
\begin{align}
    p(\by_i,\phi_i|\cdot) 
    &= p(\phi_i|\cdot)p(\by_i|\cdot)\\
    &=\frac{\left[\sum_{j=1}^c f_j(\bx_i)\right]^{n_i}}{\Gamma(n_i)}\phi_i^{n_i-1}\exp\left\{-\phi_i\sum_{j=1}^c f_j(\bx_i)\right\} \frac{n_i!}{y_{i1}!y_{i2}!...y_{ic}!}\left[ \frac{\prod_{j=1}^cf_j(\bx_i)^{y_{ij}}}{[ \sum_{\ell = 1}^c f_\ell(\bx_i)]^{n_i} }  \right]     \\
    &=\frac{n_i!}{y_{i1}!y_{i2}!...y_{ic}!}\frac{ \phi_i^{n_i-1} }{\Gamma(n_i)}\exp\left\{-\phi_i\sum_{j=1}^c f_j(\bx_i)\right\} \left[ \prod_{j=1}^cf_j(\bx_i)^{y_{ij}}  \right]     \\
    &=\frac{n_i!}{y_{i1}!y_{i2}!...y_{ic}!}\frac{ \phi_i^{n_i-1} }{\Gamma(n_i)}\prod_{j=1}^c f_j(\bx_i)^{y_{ij}}  \exp\left\{ -\phi_if_j(\bx_i)\right\}. \label{eq:joint}
\end{align}
The product in Equation \ref{eq:joint} means there are new avenues to computational efficiency because the posterior parameters of $f_j(\bx_i)$ need not depend on other $f_k$, $k \ne j$.

\subsection{Our Proposed Method}
\label{sec:ours}
We adapt this data augmentation idea from \cite{Murray} for use in standard multinomial logistic regression, i.e. we let $f_j(\bx_i) = \exp\{\bx_i^T \bbeta_j\}$. 
Thus we likewise introduce the auxiliary variables $\phi_i$ as 
\begin{equation}\label{def:phi}
    (\phi_i|\cdot) \sim \text{Gamma}\left(n_i, \sum_{k=1}^C \exp(\bx_i^T \bbeta_k)\right). 
\end{equation}

Now, conditioning the full joint likelihood on the data and the set of $\phi_i$, we get a posterior distribution of $\bbeta_j$ that is no longer proportional %in $\bbeta_j$ 
to any of the other categories' coefficients $\bbeta_k$ (for $k \neq j$), as seen in Equation \ref{conditional:beta}.
\begin{align}
    p(\bbeta_j|\cdot)
    &\propto  
    \left[\prod_{i=1}^N   p(\by_i,\phi_i|\cdot) \right]p(\bbeta_j)\\
        &= \left[\prod_{i=1}^N \frac{n_i!}{y_{i1}!y_{i2}!...y_{ic}!}\frac{ \phi_i^{n_i-1} }{\Gamma(n_i)}\prod_{j=1}^c \exp\left\{y_{ij}\bx_i^T\bbeta_j  -\phi_i \exp\left(\bx_i^T\bbeta_j\right)\right\}  \right]  p(\bbeta_j)  \\
    &
    \propto
    \exp\left\{\sum_{i=1}^N y_{ij}\bx_i^T\bbeta_j  -\phi_i \exp\left(\bx_i^T\bbeta_j\right)\right\}     p(\bbeta_j).  \label{conditional:beta}
\end{align}

Thus, sampling from the model has two main steps: 
\begin{enumerate}
    \item For each $i = 1,...,N$, sample $\phi_i$ from the distribution in Equation \ref{def:phi} conditioning on the data and the current draws/values of $\bbeta_j$, $j=1,...,C$. 
    \item For each $j=1,...,C$, sample a new $\bbeta_j$, conditioning on the data and the current draws/values of $\phi_i$, $i=1,...,N$ as in Equation \ref{conditional:beta}. 
\end{enumerate}

Above, we do not specify how to sample $\bbeta_j$. Unlike most other data augmentation techniques discussed in Section \ref{sec:intro}, the posterior distribution in Equation \ref{conditional:beta} does not have an obvious conjugate prior to assume for $p(\bbeta_j)$, and as such the choice is up to the practitioner.  As these are non-conjugate, we are not proposing a full Gibbs sampler. Instead, we  demonstrate two useful options for sampling the posteriors of $\bbeta_j$.  Then in our analyses in Section \ref{sec:sim} and \ref{sec:data}, we choose standard normal priors for all coefficients.

% \subsubsection{Elliptical Slice Sampling}
First, we use elliptical slice sampling from \cite{MurrayAdamsMackay2010}. This method of slice sampling requires a $P$-dimensional multivariate normal prior on each vector $\bbeta_j$, $j=1,...,C-1$. For each sampling step, a random draw from the prior is combined with the current state of the sampler to create an ellipse of plausible values in $\mathbb{R}^P$ space. Proposal values are taken from this ellipse and accepted/rejected based on the log likelihood.  

% \subsubsection{Adaptive Metropolis-Hastings}
Second, like \cite{johndrow2019mcmc}, we use a sequence of univariate adaptive Metropolis samplers for each coefficient in each of the $\bbeta_j$ vectors. This simple approach preserves flexibility and can generalize to many scenarios. In contrast, a multivariate normal proposal distribution would require a case-specific covariance matrix to sample efficiently, such that there are $C-1$ different $P\times P$ covariance matrices to tune. Instead, we simply have $(C-1)*P$ univariate proposal distributions, each with a single variance value to tune.  Akin to the adaptive Metropolis-within-Gibbs method of \cite{roberts2009examples}, we tune these using the approach of \cite{Fellingham2018PredictingHR}, where the proposal variances are adjusted during the burn-in phase of MCMC such that the acceptance rate is in an acceptable range, then the proposal variances are all fixed for the actual saved MCMC samples. Further detail is in Appendix \ref{app:tuning}.

\section{Simulation Studies} 
\label{sec:sim}
In order to compare the performance of these algorithms for different numbers of categories (C), samples (N), and covariates (P), we conduct a simulation study using the following data-generating process.  
\begin{align*}
    x_{ip} &\sim \text{Normal}(0,1), && i=1,...,N; p=1,...,P\\
    \beta_{jp} &\sim \text{Uniform}(0,1), && j=1,...,(C-1); p=0,...,P\\
    \bm{p}_i &= \text{softmax}(\bx_i B), && i=1,...,N\\
    \by_i &\sim \text{Multinomial}(n_i,\bm{p}_i), &&  i=1,...,N
\end{align*}
Given a fixed number of categories $C$, $N$ vectors $\bx_i \in \mathbb{R}^P$ were independently simulated from the univariate standard normal distribution, such that each subject $i$ has P predictors plus an intercept $x_{i0}=1$.  Likewise, $(C-1)$ length-$(P+1)$ vectors $\bbeta_j$ of coefficients were generated, but from the univariate standard uniform distribution, which is used instead of the standard normal as it allows the simulated data to have fairly balanced categories. The last category is considered the baseline, such that $\bbeta_C = 0$.  
For simplicity of notation, let these coefficients be combined into a $(P+1) \times C$ matrix $B$. To generate categorical probabilities,  we set $\bm{p}_i = \text{softmax}(\bx_i  B)$, then with $n_i=1$, randomly draw the categorical variable $\by_i$ according to probabilities $\bm{p}_i$ and pre-specified category quotas. These quotas will allow us to control how balanced or imbalanced the simulated datasets are.

Both of our proposed samplers, i.e. data augmentation with either adaptive Metropolis-Hastings (DA+AMH) or elliptical slice sampling (DA+eSS)\footnote{Note that we use ESS to denote effective sample size, while eSS denotes elliptical slice sampling.}, are built into an R package with C++ code using RCPP \citep{Rcpp}, available on Github \citep{BayesMultiLogit}. 
We compare our methods to adaptive Metropolis-Hastings (AMH), P\'olya-Gamma (PG), and Ultimate P\'olya-Gamma (UPG). 
AMH is implemented using the technique from \cite{Fellingham2018PredictingHR} built in C++ but accessed through R. 
For PG, \cite{Windle2013}'s current CRAN package \texttt{BayesLogit} only contains a P\'olya-Gamma random number generator, but we reconstructed their multinomial logistic regression based on their paper and previous code available on \cite{ScottPGCode} and \cite{WindleThesisCode}. 
UPG comes from \cite{zens2023ultimate}'s \texttt{UPG} package on CRAN, \cite{upgPackage}; we note that this package is entirely in R code, and while the effective sample sizes are competitive, we will see the computation speed is not.

 All computation was performed on a server with an Intel Xeon Gold 6438Y+ CPU, with each process limited to a single thread as this work analyzes algorithm performance and not the processor's ability to process calculations in parallel, as we have found that this varies greatly across machines.  Each method runs for 6,000 MCMC iterations, and 3,000 iterations are discarded as burn-in.

\subsection{Balanced Categories}
\label{sec:balanced}

We fix $N=1000$ subjects with one observation each, $P=10$ covariates, and vary $C \in \{5,10,15,20,...,95,100\}$. For each $C$ we generate datasets with equal per-class quotas ($N/C$ observations per category, rounded up or down to the nearest integer if necessary) to isolate the effect of $C$ on computation. 20 simulation replicates were run for each scenario, and  scenarios ``timed out" and were dropped if they could not complete 6000 samples within 5 minutes across all 20 replicates. Thus we see that UPG only goes up to 25 categories, while PG goes up to 35. Timeouts are interpreted as evidence of poor scalability under the same resource constraints.

\begin{figure}
    \centering
        \includegraphics[width=0.9\linewidth]{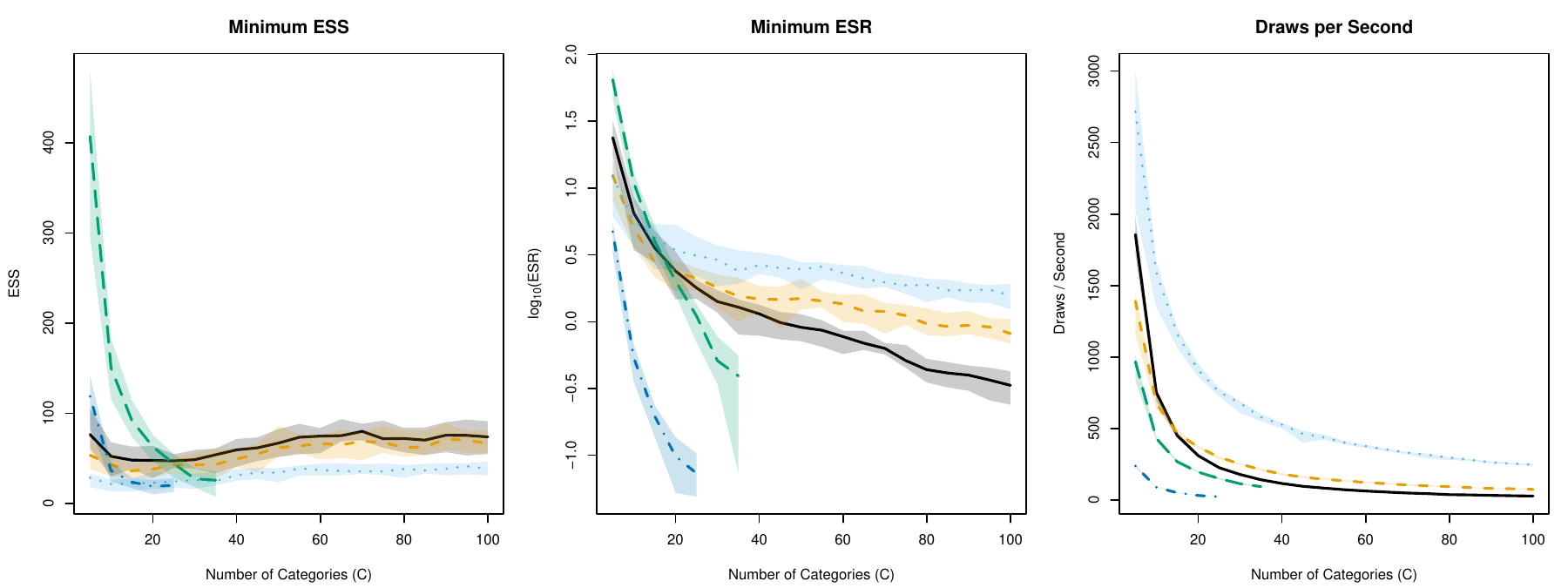}
    \includegraphics[width=0.9\linewidth]{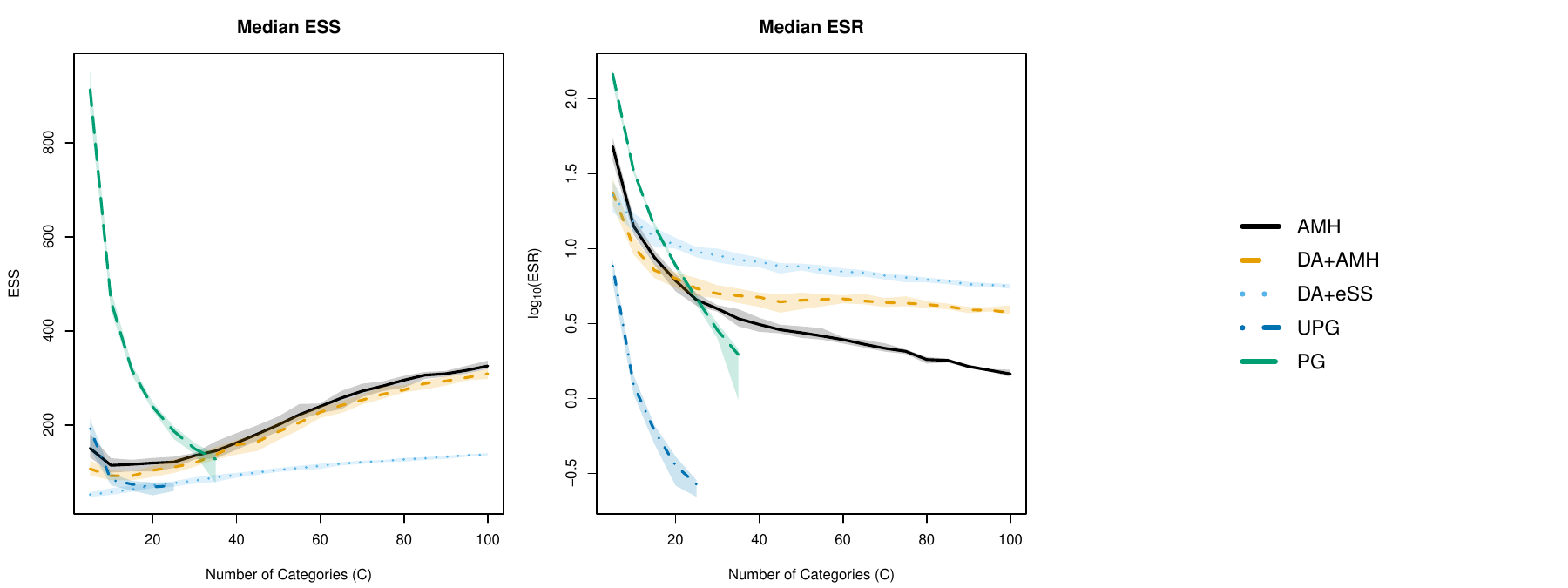}
    \caption{Comparison of posterior sampling performance metrics: MCMC iterations per second, effective sample size (ESS), and effective sampling rate (ESR). 
    The minimum ESS and ESR are the minimums over all 10*(C-1) parameters, for each simulation replicate; likewise the median ESS and ESR are the medians over all 10*(C-1) parameters. 
    The plotted lines are the median values across 20 simulation replicates, and the shaded regions cover the 5th to 95th percentiles. For each simulation replicate, each method ran for 6000 MCMC samples with 3000 discarded as burn-in.}
    \label{fig:main}
\end{figure}

Figure \ref{fig:main} presents the simulation results, where the lines represent the median value across all simulations, while the shaded regions highlight the 5th and 95th percentiles of the measure of interest. The right panel shows the number of sampling iterations per second (effectively the inverse of computation time, such that higher values are preferred). We see that DA+eSS is faster than the bulk of the methods, while UPG is slower than the others. 

The remaining panels of Figure \ref{fig:main} show the minimum or median effective sampling rate of the methods for different scenarios. The effective sampling rate (ESR) is defined as the effective sample size (ESS) divided by the number of seconds required for computation. 
Because ESR values can span orders of magnitude as $C$ increases, we plot log10(ESR).  
As each model is estimating $(P+1)(C-1)$ regression coefficient parameters, we look at the median (or minimum) ESR across all parameters.  We see that for larger numbers of categories,  DA+eSS and DA+AMH yield the highest ESRs for both the hard-to-sample parameters (seen by the minimum ESR) and the ordinary parameters (seen by the median ESR). For DA+eSS in particular, we see in the left panels that it has a rather low ESS, but it makes up for this with the fastest runtime of all methods.  At $C=100$ categories, DA+AMH has minimum ESR about 2.6 times that of AMH, and the same holds for the median ESR.  Furthermore, DA+eSS has minimum ESR about 5 times that of AMH, and median ESR about 3.8 times that of AMH. 

However, we see that for a small number of categories, the traditional PG produces very high ESSs, yielding high ESR when there are fewer than 15 or 20 categories. For instance, again using AMH's ESR as a baseline at $C=10$ categories, DA+AMH has smaller ESRs at about 75\% of that of AMH. DA+eSS meanwhile performs similarly to AMH. However, PG's ESRs are about twice that of AMH. 

As UPG is written in R, not C++, we focus on comparing its ESS. For a small number of categories, its ESS is on par with AMH and DA+AMH, though all are less effective than PG. Its performance drops quickly though as $C$ increases and by $C=20$ its ESS is similar to DA+eSS.

\subsection{Imbalanced Categories}

To reflect common real-world imbalance, we generate datasets with per-class quota $q$ for the first $(C-1)$ categories and allocate the remaining $(N-(C-1)q)$ subjects (with one observation per subject) to the final category as the majority/dominant class. We examine $q \in \{10,20,30,40,50\}$ at fixed $C=20$, $N=1000$, $P=10$ to trace how varying degrees of imbalance affect sampler efficiency. 
$q = 50$ subjects per category would be a balanced sample, while decreasing the minority classes' sizes $q=10$ leads to a 81:1 class ratio. 

\begin{figure}
    \centering
        \includegraphics[width=0.9\linewidth]{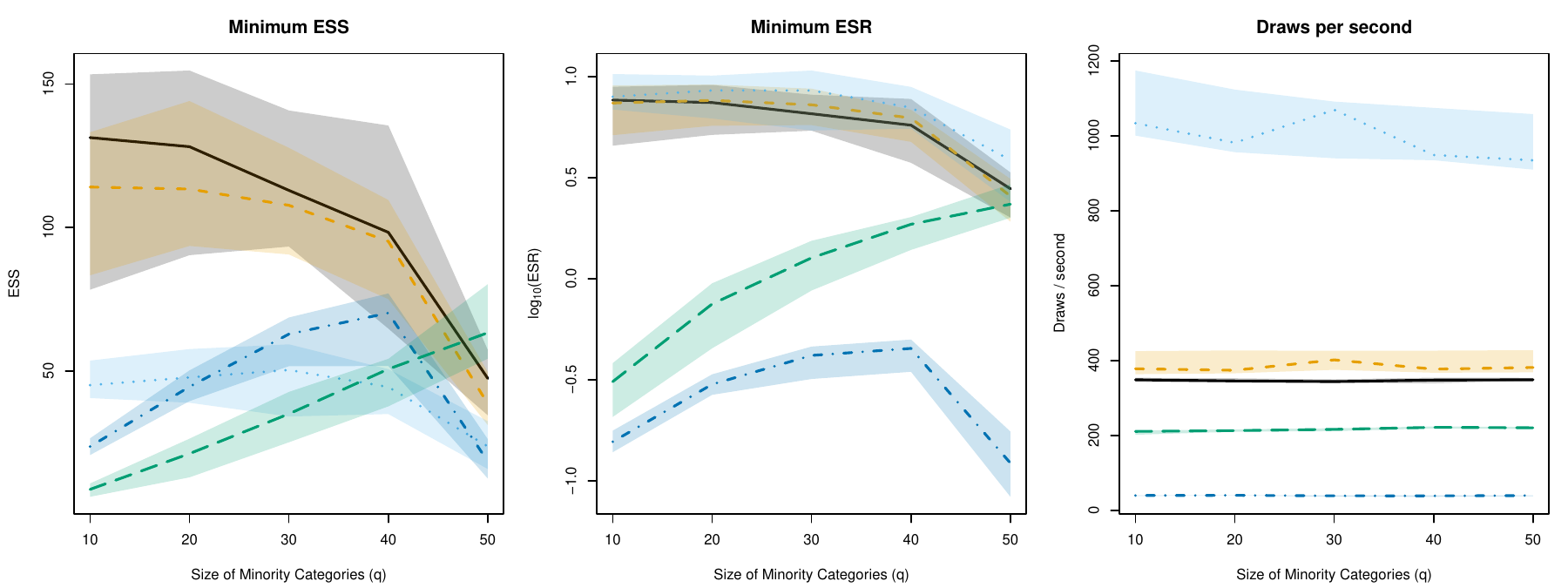}
    \includegraphics[width=0.9\linewidth]{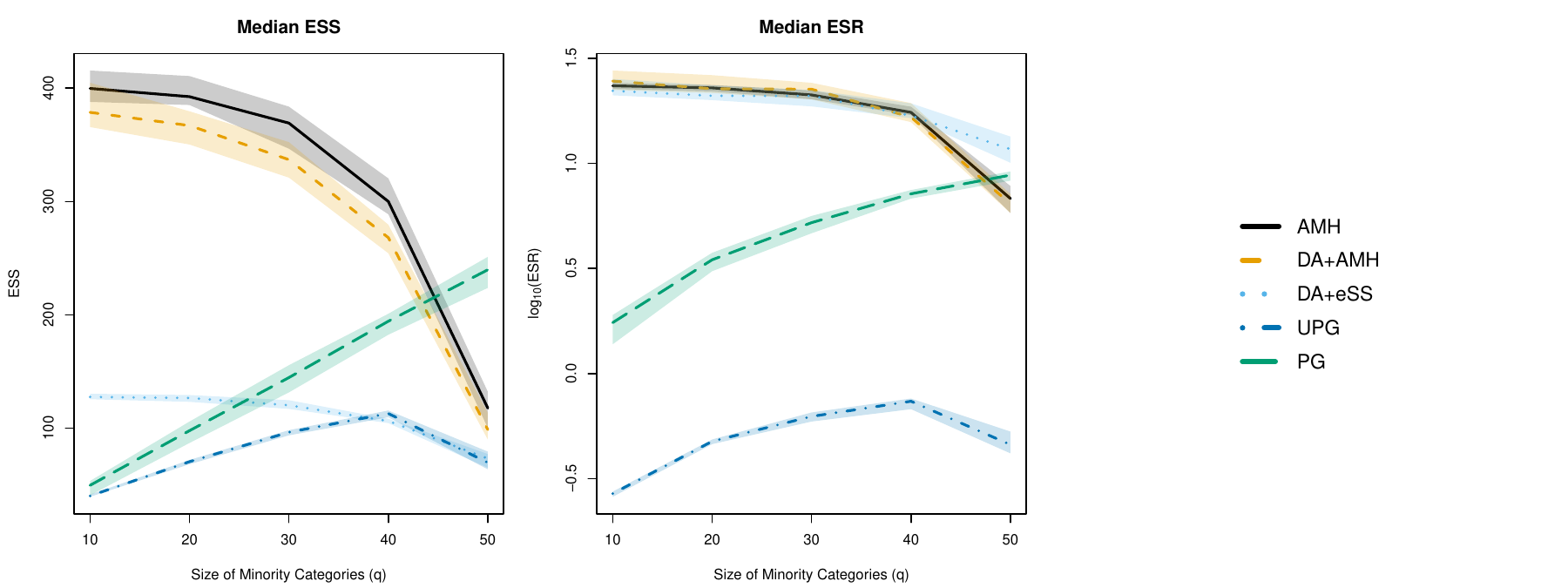}
    \caption{Comparison of posterior sampling performance metrics: MCMC iterations per second, effective sample size (ESS), and effective sampling rate (ESR).     The minimum (or median) ESS and ESR are the minimums over all $(P+1)*(C-1)=209$ parameters.  The plotted lines are the median values across 20 simulation replicates, and the shaded regions cover the 5th to 95th percentiles. For each simulation replicate, each method ran for 6000 MCMC samples with 3000 discarded as burn-in.}
    \label{fig:imbalanced}
\end{figure}

The right panel of Figure \ref{fig:imbalanced} shows that MCMC runtime does not depend on the degree of imbalance in the data. We do, however, see that ESRs for AMH, DA+AMH, and DA+eSS are similar to each other for any given value of $q$. This suggests that all three methods are equally effective at addressing imbalance across categories.  We also see the peculiar result that ESS and ESR tend to increase as the sample becomes more unbalanced. Having fewer data points per category naturally makes computation for some posteriors faster. UPG and PG meanwhile tend to be less efficient as the imbalance increases (small $q$).

\cite{zens2023ultimate} finds similar results when examining $C=3$ categories. They examine the the spectral density of the posterior chain of the model intercept, and as such our ESS results are most comparable. Among the methods in their simulation study are AMH, UPG and PG. They find that, in terms of inefficiency factors, PG is the best method when balanced, though AMH is best when heavily imbalanced, while UPG performs reasonably in both cases. Similarly, we see in Figure \ref{fig:imbalanced} that PG's ESS is highest when categories are balanced, and that AMH's ESS is highest when the categories are very imbalanced. We also see that UPG's ESS is preferable to DA+eSS and PG when imbalance is moderate ($q =30,40$). 
Both our study and theirs find that AMH's performance can actually improve for some versions of imbalance, and that UPG has a nonmonotonic relationship between degree of imbalance and sampling efficiency \citep[Fig.~3]{zens2023ultimate}.

\section{Real Data Analysis}
\label{sec:data}
We illustrate the methods on the UCI Letter Recognition dataset \citep{letter_recognition_59}. With $N = 20000$ subjects and $C= 26$, the categories are basically balanced, with the smallest at 734 observations and largest 813. We standardize the $P=16$ predictors and fit a multinomial logistic regression model with the final category as the baseline. For each method we run 6,000 iterations (3,000 burn-in) and summarize ESS and ESR across all $(P+1)(C-1) = 425$ coefficients. 

 Table \ref{tab:uci} contains the runtimes in minutes, the effective sample sizes (ESS), and effective sampling rates (ESR, the ESS per minute) for 3000 posterior samples after 3000 burn-in iterations. DA+eSS finished 6000 samples in 6.4 seconds, while AMH and DA+AMH needed about 2-3 times as long, taking 14.9 and 17.7 seconds respectively. UPG took much longer, but again, is coded in R and not C++. PG is excluded as it timed out after 24 hours.

% latex table generated in R 4.5.2 by xtable 1.8-4 package
% Tue Jan 27 10:24:10 2026
\begin{table}[bt]
\centering
\begin{tabular}{lr|rr|rr}
  \hline
   &  Runtime& \multicolumn{2}{c|}{ESS} & \multicolumn{2}{c}{ESR} \\
  \cline{3-4} \cline{5-6}
 Method &  (Minutes) & Minimum & Median & Minimum & Median \\  \hline
 AMH & 14.9 & 5.9 & 42.3 & 0.4 & 2.8 \\ 
 DA+AMH & 17.7 & 3.1 & 24.3 & 0.2 & 1.4 \\ 
 DA+eSS & 6.4 & 1.4 & 8.3 & 0.2 & 1.3 \\ 
 UPG & 147.4 & 2.9 & 15.4 & 0.0 & 0.1 \\ 
   \hline
\end{tabular}
    \caption{Posterior sampling performance metrics for fitting the letter recognition data. Each method was used to draw 6000 MCMC samples with 3000 discarded as burn-in. Total runtime is measured in minutes. The minimums and medians reported are the respective values  over all 425 estimated parameters, both for the effective sample size (ESS) and effective sampling rate (ESR) defined as effective samples per minute.}
\label{tab:uci}
\end{table}

 The ESS for every sample is quite low, indicating that, in practice, many more iterations would be needed.  Like in our simulations, DA+eSS has a low ESS as the tradeoff to fast computation speed, with UPG, DA+AMH, and AMH having two, three, and five times the effective samples respectively. Most importantly, we find that AMH's ESR is about double that of DA+AMH and DA+eSS. 

 In Section \ref{sec:balanced}, we found that DA+AMH, and DA+eSS slightly outperform AMH similarly when C=25, N=1000, P=10 (DA+AMH ESR's are 1.2x of AMH, while DA+eSS are about double).   As the number of categories (C) and covariates (P) in this data set are similar to the aforementioned simulation, the improved performance of AMH relative to DA+eSS and DA+AMH is likely attributable to the 20-fold increase in sample size. 

\section{Discussion}
\label{sec:discussion}
In this paper we present two contributions to the literature on Bayesian multinomial logistic regression. First, we propose a data-augmentation approach that improves computational efficiency in cases with many categories. Our data augmentation approach can be coupled with many traditional sampling methods, such as Metropolis-Hastings or slice sampling, and have competitive or superior effective sampling rates compared to other leading methods in the field when working with many categories. 

Second, we analyze cases that are infrequently explored in the current literature. Most works on multinomial logistic regression look at scenarios with three categories and only one observation per subject, i.e. categorical likelihoods. In our simulation studies, we demonstrate the efficiency of the various methods across the number of categories as well as the degree of imbalance.  
These methods include our proposed data-augmentation methods, the P\'olya-Gamma augmentation of \cite{Windle2013}, adaptive Metropolis-Hastings inspired by \cite{johndrow2019mcmc}, and the Ultimate P\'olya-Gamma method of \cite{zens2023ultimate}.

There are several possible extensions to our analysis. There are other adaptive Metropolis techniques we could have used, such as the adaptive Metropolis of \cite{haario2001adaptive} used in \cite{johndrow2019mcmc}. 
Another potential direction would be exploring truly Multinomial situations, i.e. $n_i \geq 1$, however, in previous work we found that results were similar to what we have shown here, such that they are not included in this paper.  

We have implemented these ideas and techniques into an R package available on Github \citep{BayesMultiLogit}. Another important extension of this work could be to parallelize the sampling of the regression coefficients and/or the augmented variables, which we currently do not employ. The augmented variables could be drawn in parallel across subjects, but as each is simply a draw from a gamma distribution, there is not high potential for computational savings. Still, the coefficients can be drawn in parallel across categories, and as this is the more complicated operation of the two, and as the number of categories is often comparable to a machine's number of cores, this is an intriguing prospect. However, three different attempts to parallelize these draws have shown that the computational overhead of spinning up parallel processes is not mitigated by the computational savings of coefficient sampling done in parallel.

\newpage
\bibliographystyle{apalike}
\bibliography{References}

\appendix
\section{Appendix}
\label{app:tuning}
Our univariate Metropolis samplers of the $\bbeta_j$ coefficients are set up to automatically tune the proposal distribution parameters during the burn in iterations, and this proceeds as follows. 

Let $\beta_{jp}^{(t-1)}$ be the coefficient in category $j=1,...,C-1$ for predictor $p=0,...,P$, drawn at MCMC iteration $t-1$. The proposal distribution for generating $\beta_{jp}^{(t)}$ is $q(\beta_{jp}^{(t-1)}) = \beta_{jp}^{(t-1)} + W$ where $W \sim N(0, \sigma_{jp}^2)$. While using univariate samplers mean that there is no covariance matrix to tune, there is a variance term $\sigma_{jp}^2$ to tune for each coefficient's proposal distribution, $(P+1)(C-1)$ in total. While these can be pre-specified by the user, we have automated their tuning during the burn-in phase of the sampler, akin to \cite{Fellingham2018PredictingHR}.

During the burn-in phase of MCMC, the value of $\sigma_{jp}$ is adjusted if the acceptance rate is outside of an acceptable range (20 - 40\%). Specifically, for some predetermined number of MCMC iterations $\eta$, which is much less than the number of burn-in iterations, we check the number of values accepted during those $\eta$ iterations. If there were more than $0.4\eta$ new values accepted in these $\eta$ iterations, we double the value of $\sigma_{jp}$. If there were fewer than $0.2\eta$ new values accepted, we reduce $\sigma_{jp}$ by 10\%, i.e. set its value to $0.9\sigma_{jp}$. In this way, as long as the initially provided values of both  $\eta$  and $\sigma_{jp}$ are reasonable, the $\sigma_{jp}$ will be automatically tuned to values that produce an acceptance/rejection rate in the desired range. For example, if performing 2000 burn-in iterations, then $\eta=100$ provides $2000/\eta =20$  windows with which to tune $\sigma_{jp}$ for all $j,p$.

\end{document}